# THE APPLICATION OF DATA MINING TECHNIQUES TO SUPPORT CUSTOMER RELATIONSHIP MANAGEMENT: THE CASE OF ETHIOPIAN REVENUE AND CUSTOMS AUTHORITY

**Belete Biazen Bezabeh\***

Bahir Dar University, Bahir Dar Institute of Technology, Bahir Dar, Ethiopia
Corresponding author\*, e-mail:  beleb2009@gmail.com / beleteb@bdu.edu.et

***Abstract-*** **The application of data mining technique has been widely applied in different business areas such as health, education and finance for the purpose of data analysis and then to support and maximizes the organizations' customer satisfaction in an effort to increase loyalty and retain customers' business over their lifetimes . The researchers' primary objective, in this paper is to classify customers based on their common attributes since customer grouping is the main part of customer relationship management. In this study, different characteristics of the ERCA customers' data were collected from the customs database called ASYCUDA. Once the customers' data were collected, the necessary data preparation steps were conducted on it and finally a dataset consisting of 46748 records was attained. The classification modeling was built by using J48 decision tree and multilayerperceptron ANN algorithms with 10-fold cross-validation and splitting (70% training and 30% testing) techniques. Among these models, a model which was built using J48 decision tree algorithm with default 10-fold cross-validation outperforms 99.95% of overall accuracy rate; while the classification accuracy of ANN is 99.71%. So decision tree has better accuracy than ANN for classifying ERCA customers' data.**

## 1. INTRODUCTION

Customer relationship management (CRM) has become one of the strategies of an organization for sustained competitive advantage. CRM in its broadest sense simply means managing all customer interaction (Trappey et al. 2009). The new millennium is in the middle of explosive change witnessing rapidly changing market conditions, volatile equity markets, reconstructed value chains and new global competitors (Kumar and Solanki 2010). Customers themselves are changing, and consider natural customer loyalty which is a thing of the past (Suresh 2002). CRM includes all measures for understanding the customers and for exploiting this knowledge to design and implement marketing activities, align production and coordinate the supply-chain (Srivastava 2002*)*.

Customer segmentation is the grouping of customers into different groups based on their common attributes and it is the main part of CRM (Verhoef 2003). Segmentation requires the collection, organization and analysis of customer data. With proper segmentations of a customer's data it is possible to identify the reliability/loyalty of customers so as to increase the revenue of the organization. CRM creates interaction of customers with the organization by using information technology (IT). Moreover, identifying customer's need/interest better and treating them accordingly can increase their life time (Verhoef 2003).

## 2. STATEMENT OF THE PROBLEM

Currently, the ERCA is using statistical analysis and assessment techniques to identify potential and low valued customers. The Authority check whether the customers discharge their responsibility or not in the revenues database and at the same time they cross check the customs database whether the items are imported/exported with paying the required tax by using assessment and statically analysis techniques. However, these techniques are not effective and efficient and took a considerable amount of time to treat customers according to their characteristics. These techniques are also less efficient to control taxpayers who fail to declare their actual income in order to reduce their tax bill and the federal government's revenue.

## 3. METHODOLOGY

This study generally follows both quantitative and qualitative methods. The quantitative method is used to collect and analyze customers' data. On the other hand, the qualitative aspect is used to understand the business operation by making a close relationship with a domain experts and the responsible body such as the Database Administrator of the organization. On this study a six step KDD process model developed by Cios et al. (2000) is considered because this model combines both academics and industry aspects. In this study WEKA as a tool is used for classifications.

## 4. DATA MINING PROCESS

Data mining process, as depicted in figure 1 below, is a step in KDD process which consists of methods that produce useful patterns or models from the data (Nasereddin 2009).

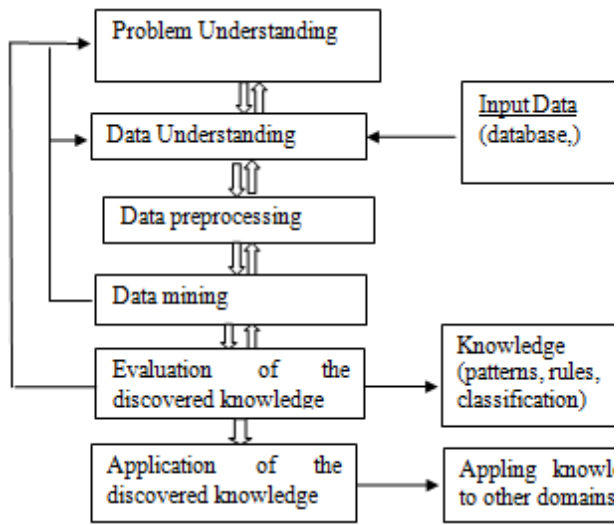

**Figure 1. The six-step Cios et al. (2000) KDD process model**

In data mining process there are two possibilities; in some cases when the problem is known and correct data is available as well, there is an attempt to find the models or tools which will be used. On the other hand, some problems might occur because of duplicate, missing, incorrect, outlier values and there is a need to make some statistical methods.

## 5. EXPERIMENTATION

Classification is a learning model in data mining techniques which aims at building a model to predict future customer behaviors through classifying database records into a number of predefined classes based on certain criteria. In this study, most common classification techniques, such as decision tree and neural network classification techniques were tested; for decision tree J48 algorithm and for neural network Multilayer-Perceptron algorithm were investigated.

The output of the selected clustering model was fed to the J48 decision tree algorithm. Here the cluster index is used as the dependant variable, whereas the remaining all attributes which are selected for the cluster model building, are fed as independent variables. The J48 decision tree provided a descriptive classification model of the clusters, thus enabling exploration and detection of the characteristic of each cluster. During the generation of a classification model, both options, the 10-fold cross-validation and percentage split (with 70% train and 30% test), were investigated. When using the J48 decision tree algorithm with 10-fold cross-validation default parameter value, the output of the tree consisted of 91 nodes and 54 leaves. The output of the confusion matrix for this learning algorithm looks as follows.

| Actual | Predicted | | | | | Total | Accuracy Rate |
|---|---|---|---|---|---|---|---|
| | Cluster1 | Cluster 2 | Cluster 3 | Cluster 4 | Cluster 5 | | |
| Cluster 1 | 17633 | 0 | 1 | 0 | 6 | 17640 | 99.96% |
| Cluster 2 | 0 | 9209 | 1 | 0 | 0 | 9210 | 99.98% |
| Cluster 3 | 0 | 0 | 11742 | 0 | 4 | 11746 | 99.96% |
| Cluster 4 | 0 | 2 | 0 | 1484 | 0 | 1486 | 99.96% |
| Cluster 5 | 3 | 0 | 3 | 0 | 6660 | 6666 | 99.90% |
| Total | 17636 | 9211 | 11747 | 1484 | 6670 | 46748 | 99.95% |

**Table 1. J48 decision tree algorithm with 10-fold cross-validation default parameter value**

The output of the confusion matrix shows that, in this experiment from the total 46748 amounts of data, 46728 (99.95%) of the records were correctly classified, while the remaining 20 (0.05%) of the records were incorrectly classified.

Table 4.18 clearly shows that, from 17640 records, 17633 (99.96%) of the records were correctly classified as Cluster One (low-value customers), while the remaining 1 record was incorrectly classified as Cluster Three, and the other 6 were incorrectly classified as Cluster Five. From 9210 records, 9209 (99.98%) of the records were correctly classified as Cluster Two (high-value customers), whereas the remaining 1 record was incorrectly classified as Cluster Three. Out of 11746 records, 11742 (99.96%) records were correctly classified as Cluster Three, while the remaining 4 records were incorrectly classified as Cluster Five. From 1486 records, 1484 (99.96%) records were correctly classified as Cluster Four, whereas the remaining 2 records were misclassified as Cluster Two (high-value customer). Out of 6666 records, 6660 (99.90%) records were correctly classified as Cluster Five; while the remaining 3 records were misclassified as Cluster One (low-value customers) and the other 3 records were also incorrectly classified as Cluster Three.

Although the above experiment had good accuracy, the researcher tried to find the best classification with small tree size; because small tree size decision tree classifications are easier to generate rules. So, the researcher tried various experiments by changing the default value of the J48 decision tree 10-fold cross-validation parameter values. The default values of 10-fold cross-validation are the minimum number of instances per leaf (minNumObj) with 2 and the confidence factor used for pruning (confidenceFactor) with 0.25. The different values of minNumObj with 5, 10, 15, 20, 25 were tested and the output of the confusion matrix for minNumObj=25 displayed in the following table.

| Actual | Predicted | | | | | Total | Accuracy Rate |
|---|---|---|---|---|---|---|---|
| | Cluster1 | Cluster 2 | Cluster 3 | Cluster 4 | Cluster 5 | | |
| Cluster 1 | 17583 | 0 | 29 | 20 | 8 | 17640 | 99.67% |
| Cluster 2 | 0 | 9209 | 1 | 0 | 0 | 9210 | 99.98% |
| Cluster 3 | 0 | 6 | 11727 | 0 | 13 | 11746 | 99.83% |
| Cluster 4 | 5 | 2 | 0 | 1479 | 0 | 1486 | 99.52% |
| Cluster 5 | 12 | 0 | 16 | 14 | 6624 | 6666 | 99.36% |
| Total | 17600 | 9217 | 11773 | 1513 | 6645 | 46748 | 99.73% |

**Table 2 J48 decision tree algorithm with minNumObj=25 and confidenceFactor=0.25**

| Actual | Predicted | | | | | Total | Accuracy Rate |
|---|---|---|---|---|---|---|---|
| | Cluster1 | Cluster 2 | Cluster 3 | Cluster 4 | Cluster 5 | | |
| Cluster 1 | 5264 | 0 | 10 | 7 | 0 | 5281 | 99.67% |
| Cluster 2 | 0 | 2748 | 0 | 0 | 0 | 2748 | 100% |
| Cluster 3 | 0 | 6 | 3559 | 0 | 8 | 3573 | 99.60% |
| Cluster 4 | 2 | 0 | 0 | 437 | 0 | 439 | 99.54% |
| Cluster 5 | 6 | 0 | 4 | 6 | 1967 | 1983 | 99.19% |
| Total | 5272 | 2754 | 3573 | 450 | 1975 | 14024 | 99.65% |

**Table 3 Summary of the confusion matrix with default parameters value and 70 % for training and 30 % for testing dataset**

Compared with the previous experiments, both 10-fold cross-validation experiments, this experiment had the least overall accuracy rate. Even in the individual cluster classification, except for the second cluster, the 10-fold cross-validation had better classification accuracy than the percentage split experimentation. Hence, from all of the above experimentations, the first model, 10-fold cross-validation with the default parameter values, was selected because this model had better accuracy both in the overall and individual cluster classification.

### 5.1 ANN CLASSIFICATION MODEL

The other common classification technique is the Artificial Neural Network (ANN) classification model. According to Han and Kamber (2006), ANN classification model learn very fast when the attributes' values fall in the range [-1, 1]. So, to put the attribute value in the range of [-1, 1], the researcher used the Weka normalizing preprocessing facilities. Consequently, all numeric attributes were normalized; that is their value fell in between the range of [-1, 1]. However, CTY_DSC (the origin of items) is a nominal attribute but as it is stated in Section 2.6.2.2, ANN classification model usually works only with numerical data. So, there are six distinct values in this attribute and each of them is assigned a numeric value from 1-6. After mapping the nominal value into the numeric value, the Weka preprocessing facility normalizes all values to fall in the range of [-1, 1].

Hence, the same attributes that were used to build the decision tree models, were also used in the neural net modeling .Like the J48 decision tree algorithm in this experiment also the researcher used both the 10-fold cross-validation and percentage split tests. During the experimentation various runs are performed by using the default values and changing the hidden layers, learning rate and momentum parameter values. The default hidden layers, learning rate and momentum parameters are shown in the following table.

| Parameter | Description | Default value |
|---|---|---|
| Hidden Layers | This defines the hidden layers of the neural network. This is a list of positive whole numbers. | 'a' = (attributes + classes) / 2 a=(8 (number of attributes +5 (number of clusters))/2=6.5 (7) |
| Learning Rate | The amount the weights are updated | 0.3 |
| momentum | Momentum applied to the weights during updating | 0.2 |
| 10-folds cross-validation | | |

**Table 4 Parameters and their default values of the neural network classifier**

As it is shown in Table 4.22, the multilayer perceptron ANN algorithm had good overall and individual cluster classification accuracy rate. From the total dataset (46748) records, 46613 (99.71%) records were correctly classified by this algorithm; only 135 (0.29%) records were misclassified.

For the percentage split run also better performance accuracy rate was found when hidden Layers=8, learning Rate=0.5 and momentum=0.4. For this split run like in the above J48 decision tree experiment, 70% of the records were used for training and 30% of the records were used for testing. The output of the confusion matrix for this split run is presented in the following table.

| Actual | Predicted | | | | | Total | Accuracy Rate |
|---|---|---|---|---|---|---|---|
| | Cluster1 | Cluster 2 | Cluster 3 | Cluster 4 | Cluster 5 | | |
| Cluster 1 | 5048 | 0 | 44 | 0 | 0 | 5092 | 99.04% |
| Cluster 2 | 0 | 2659 | 0 | 0 | 9 | 2668 | 99.66% |
| Cluster 3 | 0 | 2 | 3641 | 0 | 0 | 3643 | 99.95% |
| Cluster 4 | 0 | 0 | 0 | 345 | 0 | 345 | 100% |
| Cluster 5 | 0 | 0 | 0 | 0 | 2276 | 2276 | 100% |
| Total | 5048 | 2661 | 3685 | 345 | 2285 | 14024 | 99.60% |

**Table 5 Split output from Multilayer-Perceptron ANN algorithm with hidden-Layers=8 learning-Rate=0.6 and momentum=0.4**

The output of the split run shows that from the total 14024 testing dataset, 13969 (99.60%) of the records were correctly classified, while the remaining 55 (0.40%) records were misclassified. For individual cluster level classification the split run better correctly classified the medium-value (Cluster 4 and 5) customers and high-value (Cluster 2) customers. However, this run was still less efficient on classifying low-value (Cluster 1) customers.

### 5.2 COMPARISON OF DECISION TREE AND NEURAL NETWORK MODELS

So far two classification models were tested; the next step was comparing the above two classification models, which were the J48 decision tree model and Multilayer Perceptron ANN classification model. The purpose of the comparison was to choose the best from these two algorithms, which was appropriate for the problem domain of this research, CRM.

From the above two classification models the best algorithm was selected based on the following three parameters.

Moreover, the split run had less overall accuracy rate than 10-fold cross-validation.

From the above two Multilayer-Perceptron ANN classification models, the first model built using 10-fold cross-validation with hidden-Layers=8 learning-Rate=0.5 and momentum=0.4 was selected; because this model had better accuracy rate than the second split model.

- ➢ The overall classification accuracy rate
- ➢ The model accuracy in classifying high value customers
- ➢ The model accuracy in classifying low value customers

So, based on the above criteria the algorithm, which had high overall accuracy rate and high accuracy in correctly classifying high value and low value customers in their clusters were selected. Consequently, the comparison of the decision tree and ANN models are described as follows.

| 10-fold classification model | Overall accuracy (46748 records) | | High-value customers (Cluster 2) accuracy | | Low-value customers (Cluster 1) accuracy | |
|---|---|---|---|---|---|---|
| | Correctly classified | Incorrectly classified | Correctly classified | Incorrectly classified | Correctly classified | Incorrectly classified |

| Decision tree model | 46728 99.95% | 20 0.05% | 9209 99.98% | 1 0.02% | 17633 99.96% | 7 0.04% |
| Neural network model | 46613 99.71% | 135 0.29% | 8834 99.47% | 47 0.53% | 16744 99.70% | 48 0.30% |

**Table 6 Summary of the accuracy level of the decision tree and neural net classification models**

## 6. CONCLUSION

The classification models were built with J48 decision tree and multilayer perceptron neural net algorithms. From these two classification models the best classification model was selected by comparing the overall accuracy, accuracy in classifying high value customers and accuracy in classifying low value customers.

The model which was developed with J48 decision tree algorithm had 99.95% overall accuracy rate, whereas the multilayer perceptron neural net algorithm had 99.71% of overall accuracy. For classifying high value customers, the decision tree algorithm had 99.98% of accuracy, while the neural net algorithm had 99.47% of accuracy. Moreover, in classifying low value customers the decision tree algorithm had 99.96% of accuracy, while neural net algorithm had 99.70% of accuracy. Since, the decision tree model had scored better performance in all these evaluation parameters; it is the researcher's belief that decision tree classification model has an appropriate technique for this research on CRM. In general, the results from this study were encouraging. It was possible to segment customers' data using data mining techniques that made business sense. To this effect, related literature on data mining techniques, CRM and customer segmentation was reviewed.

As a future research direction the present work can be further investigated by increasing the number of records and adjusting the default parameter values of multilayer perceptron of ANN algorithm.